\documentclass{article}

\usepackage{arxiv}

\usepackage[utf8]{inputenc} 
\usepackage[T1]{fontenc}    
\usepackage[colorlinks=true, citecolor=blue, linkcolor=blue, urlcolor=blue]{hyperref}
\usepackage{url}            
\usepackage{booktabs}       
\usepackage{amsfonts}       
\usepackage{nicefrac}       
\usepackage{amsmath}
\usepackage{microtype}      
\usepackage{cleveref}       
\usepackage{lipsum}
\usepackage{array}
\usepackage{graphicx} 
\usepackage{subcaption}
\usepackage[round]{natbib}
\usepackage{doi}
\usepackage{xcolor}

\title{Sampled datasets risk substantial bias in the identification of political polarization on social media}

\newif\ifuniqueAffiliation

\ifuniqueAffiliation 
\author{ \href{https://orcid.org/0000-0000-0000-0000}{\includegraphics[scale=0.06]{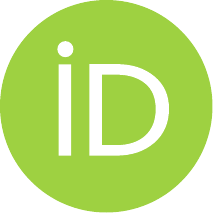}\hspace{1mm}David S.~Hippocampus}\thanks{Use footnote for providing further
		information about author (webpage, alternative
		address)---\emph{not} for acknowledging funding agencies.} \\
	Department of Computer Science\\
	Cranberry-Lemon University\\
	Pittsburgh, PA 15213 \\
	\texttt{hippo@cs.cranberry-lemon.edu} \\
	\And
	\href{https://orcid.org/0000-0000-0000-0000}{\includegraphics[scale=0.06]{figures/orcid.pdf}\hspace{1mm}Elias D.~Striatum} \\
	Department of Electrical Engineering\\
	Mount-Sheikh University\\
	Santa Narimana, Levand \\
	\texttt{stariate@ee.mount-sheikh.edu} \\
}
\else
\usepackage{authblk}

\newbox{\orcid}\sbox{\orcid}{\includegraphics[scale=0.06]{figures/orcid.pdf}} 
\author[1,2]{%
	\href{https://orcid.org/0000-0003-2615-0712}{\usebox{\orcid}}\hspace{1mm}Gabriele Di Bona\thanks{\texttt{gabriele.dibona.work@gmail.com}}%
}
\author[3]{%
	\href{https://orcid.org/0000-0002-1647-7300}{\usebox{\orcid}}\hspace{1mm}Emma Fraxanet\thanks{\texttt{emma.fraxanet@upf.edu}}%
}
\author[4,5]{%
	\href{https://orcid.org/0000-0002-8732-1258}{\usebox{\orcid}}\hspace{1mm}Björn Komander\thanks{\texttt{bkomander@iiia.csic.es}}%
}
\author[6,7,8]{%
	\href{https://orcid.org/0000-0000-0000-0000}{\usebox{\orcid}}\hspace{1mm}Andrea Lo Sasso\thanks{\texttt{andrea.losasso@uniba.it}}%
}
\author[9]{%
	\href{https://orcid.org/0000-0000-0000-0000}{\usebox{\orcid}}\hspace{1mm}Virginia Morini\thanks{\texttt{virginiamorini95@gmail.com}}%
}
\author[10,11,12]{%
	\href{https://orcid.org/0000-0002-9044-8348}{\usebox{\orcid}}\hspace{1mm}Antoine Vendeville\thanks{\texttt{antoinevendeville@sciencespo.fr}}%
}
\author[13]{%
	\href{https://orcid.org/0000-0002-2986-2494}{\usebox{\orcid}}\hspace{1mm}Max Falkenberg\thanks{\texttt{max.falkenberg@protonmail.com}}%
}
\author[14]{%
	\href{https://orcid.org/0000-0001-6859-0391}{\usebox{\orcid}}\hspace{1mm}Alessandro Galeazzi\thanks{\texttt{alessandro.galeazzi@unipd.it}}%
}

\affil[1]{CNRS, GEMASS, 59 rue Pouchet, F-75017, Paris, France}
\affil[2]{Sony Computer Science Laboratories Rome, Joint Initiative CREF-Sony, Centro Ricerche Enrico Fermi, Via Panisperna 89/A, I-00184, Rome, Italy}
\affil[3]{Department of Information and Communication Technologies, Universitat Pompeu Fabra, Tànger 122-140, 08018, Barcelona, Spain}
\affil[4]{IIIA-CSIC, Campus UAB, 08193 Cerdanyola, Spain}
\affil[5]{School of Computing Technologies, RMIT University}
\affil[6]{Universit\`a degli Studi di Bari Aldo Moro, Dipartimento Interateneo di Fisica, Bari, I-70125, Italy}
\affil[7]{Istituto Nazionale di Fisica Nucleare, Sezione di Bari, Bari, I-70125, Italy}
\affil[8]{Predict S.r.l., Viale Adriatico - Fiera del Levante - Pad. 105, I-70132 Bari, Italy}
\affil[9]{KDD Lab, CNR-ISTI, 56126 Pisa, Italy}
\affil[10]{m\'edialab, Sciences Po, 75007 Paris, France}
\affil[11]{Complex Systems Institute of Paris Île-de-France (ISC-PIF) CNRS, 75013 Paris, France}
\affil[12]{Learning Planet Institute, Research Unit Learning Transitions, 75004 Paris, France}
\affil[13]{Department of Mathematics, City University of London, London, United Kingdom}
\affil[14]{Department of Mathematics, University of Padova, Italy}

\fi

\hypersetup{
pdftitle={A template for the arxiv style},
pdfsubject={q-bio.NC, q-bio.QM},
pdfauthor={David S.~Hippocampus, Elias D.~Striatum},
pdfkeywords={First keyword, Second keyword, More},
}

\begin{document}
\maketitle

\begin{abstract}
     Following recent policy changes by X (Twitter) and other social media platforms, user interaction data has become increasingly difficult to access. These restrictions are impeding robust research pertaining to social and political phenomena online, which is critical due to the profound impact social media platforms may have on our societies. 
     Here, we investigate the reliability of polarization measures obtained from different samples of social media data by studying the structural polarization of the Polish political debate on Twitter over a 24-hour period. 
     First, we show that the political discussion on Twitter is only a small subset of the wider Twitter discussion. Second, we find that large samples can be representative of the whole political discussion on a platform, but small samples consistently fail to accurately reflect the true structure of polarization online. Finally, we demonstrate that keyword-based samples can be representative if keywords are selected with great care, but that poorly selected keywords can result in substantial political bias in the sampled data. Our findings demonstrate that it is not possible to measure polarization in a reliable way with small, sampled datasets, highlighting why the current lack of research data is so problematic, and providing insight into the practical implementation of the European Union's Digital Service Act which aims to improve researchers' access to social media data.  
\end{abstract}


\section{Introduction}


Recent years have seen a dramatic fall in the accessibility of online social media data. After recent changes in the policy of X (Twitter) and Reddit, user interaction data is more and more expensive to access, and many datasets remain inaccessible \citep{roozenbeek2022democratize}. Even before this recent upheaval, one could only obtain small samples of tweets, although tricks to circumvent this issue have been found \citep{pfeffer2023just}. Exhaustive sets of data were available through the so-called Firehose API, locked behind an expensive paywall, however few research teams had the resources to fund access. Other platforms' data access systems, such as Youtube, TikTok, and Last.fm \citep{corso2024we,di2022social}, often do not have well-documented APIs, and require a degree of reverse engineering \citep{mekacher2024koo,mekacher2023systemic}. Scholars have obtained direct access to data through collaborations with the companies themselves \citep{guess2023howdo,guess2023reshares,nyhan2023likeminded,gonzalezbailon2023asymmetric}, but this raises questions of bias, independence and reproducibility \citep{wagner2023independence}.  

For these reasons, scholars often rely on partial datasets to conduct research. In this study, we investigate the extent to which social media studies using small samples are representative of the same study carried out using full datasets. We are particularly interested in the measurement of polarization, as it is crucial aspect of social media and opinion research \citep{falkenberg2022growing,falkenberg2023affective,yarchi2021political,bramson2017understanding,abramowitz2008polarization}. Thus, we aim to assess how robust findings regarding polarization in online conversations are when the accessed data is only a sample of the full dataset. This is especially relevant to keyword-based data collection, which is inherently partial: for example, it is difficult to measure how much of the debate about Covid-19 we are actually covering when querying tweeets containing the words ``covid19'' or ``coronavirus''. These biases are not well documented, and thus such methods suffer from a lack of reliability that is hard to properly evaluate.

We conduct a simple case study that focuses on structural (sometimes referred to as interactional \citep{falkenberg2023affective,yarchi2021political}) polarization in the Polish political Twittersphere. We use Twitter due to its importance for politicians and journalists, and for its historic importance as a key data source in computational social science. At the time of collection, the political landscape in Poland has the benefit of exhibiting a clear bimodal structure (see Figure 2 in \citet{falkenberg2023affective}) between the centrist Platforma Obywatelska (Civic Platform) and the right-wing Prawo i Sprawiedliwość (Law and Justice), which makes it easy to interpret our results and provides a clear illustration of sampling biases. However, the issues we raise are neither platform- nor country-specific, and can easily be adapted to other contexts.

Our study could not only help future research in ensuring that reliable results can be obtained with minimal access to expensive APIs, but also be used as guidelines to design the technical requirement to implement the European Union's new Digital Services Act (DSA). Especially in the latter case, this analysis is key since the DSA requires data access requests to be ``proportionate'', but it is not clear where this threshold lies to ensure accurate results. Therefore, our research may provide insight and solutions as to the current social media data drought that we are currently experiencing. 



\section{Related work}

Methods have been proposed to reconstruct networks from partial data \citep{peixoto2024scalable,papanastasiou2023constrained,newman2018network,lokhov2016reconstructing,he2017notenough,gomezrodriguez2012inferring}. They offer a way to infer precise, local information about the topology of the network. However, they are often computationally intensive, and become more error-prone as we seek to reconstruct more finely. In a study close to ours, \citet{morstatter2021sample} compared partial data returned by the previously free API with a complete collection obtained via Firehose for the same period. They highlighted several biases in the unknown sampling procedure used by Twitter, such as an over-representation of content from the US and Europe, and a misrepresentation of hashtags, as the top 100 hashtags from streaming API exhibited only a 50\% correlation with those from the complete data. However, \citet{morstatter2021sample} do not focus on how partial sampling impacts the identification of polarized network structures. Therefore, in this paper we we compare a variety of different sampling methods, and focus on the specific problem of polarization.

Polarization is a broad concept that includes several nuanced sub-types \citep{kubin2021,yarchi2021political,bramson2017understanding}. Affective polarization is concerned with the prevalence of positive feelings towards similar-minded peers, and negative feelings towards others \citep{finkel2020political,iyengar2019origins}. Conversely, ideological polarization typically refers to the traditional gap between left and right \citep{abramowitz2008polarization,dimaggio1996have}, and is related to issue (also attitudinal) polarization concerning attitudes on specific topics, e.g.\ climate change \citep{falkenberg2023affective} or Covid-19 \citep{gruzd2022coordinated}. Distinguishing between issue and ideological polarization can be crucial, as some issues transcend the traditional left-right scale, such as the positions regarding European integration in France \citep{peralta2024multidimensional}. 

In this work, we focus on interactional (or structural) polarization, which characterizes to the fragmentation of interactions on social networks into distinct ideological communities \citep{yarchi2021political,flamino2023political,adamic2005political}. A vast literature is devoted to the study of such communities, which are related to the so-called echo chambers, that tend to foster the reinforcement of pre-existing beliefs and hinders communication with other groups \citep{cinelli2021echo,nguyen2020echo,garimella2018political,barbera2015tweeting,williams2015network,buongiovanni2022will,morini2021toward}. Although often only stated implicitly, the study of interactional polarization is arguably the most common sub-type studied on social media \citep{kubin2021}. 

To measure interactional polarization, studies often analyze network structure to detect segregated communities. The detection of echo chambers has gathered particular attention in that regard \citep{cinelli2021echo,garimella2018political,garimella2016quantifying,adamic2005political}. Another common technique is to use correspondence analysis, to project the nodes into ideological latent spaces which are often interpreted in reference to known political actors or partisan media outlets \citep{falkenberg2023affective,falkenberg2022growing,flamino2023political,peralta2024multidimensional,ramaciotti2022inferring,barbera2015birds}. Then, the distribution in this space is studied to determine the presence and the extent of polarization. Various metrics have been proposed to this end \citep{bramson2017understanding,duclos2004polarization}. These distributions are extracted to draw a parallel between network interactions and the typical uni-dimensional ideology distributions that are common in social science research and from which polarization can be measured quantitatively \citep{bramson2017understanding}.

\section{Materials and Methods}

Consider the following hypothetical situation: you are a researcher with a certain interest in studying political polarization on a specific platform. Considering data limitations and resources, you have to decide on a strategy to obtain the network from which you will obtain your ideology latent space and assess polarization. Since the goal is to recover politically relevant structures, our baseline is the full network obtained through political actor accounts, which has clearly identifiable actors and well-defined ideological scale. These networks are very informative and clearly polarized. 

The network of interactions can be obtained in different ways: seed-based (collect all follower or retweeters of a selected few influential accounts, e.g.\ politicians), keyword-based (collect all interactions using specific keywords, e.g.\ ``covid-19''), most popular content (collect all interactions around a few highly interacted posts). To construct such a network, one needs to query the API of the considered platform.
These APIs rarely return complete datasets, and the data obtained will be partial one way or another. It could be random sample of all data matching the query, a time-restricted sample such as all matching data in a 1h window, or restricted to only the first few accounts in the retweet list (blocked automatized cursor, a common problem with the TikTok API---see \citet{corso2024we}).

We extract samples of a complete 24h dataset of interactions following these methods, and assess how closely they resemble the full data in terms of interactional polarization. The following subsections detail the dataset, the sampling process, the ideology estimation methods and the polarization metrics considered. 



\subsection{Dataset}
We proceed similarly as in \citep{falkenberg2023affective}, and combine (i) a dataset containing all Twitter interactions on a 24 hour period \citep{pfeffer2023just}, with (ii)  a dataset of known elected politicians on Twitter \citep{vanvliet2020twitter}. The original data spans nine countries (Canada, France, Germany, Italy, Poland, Spain, Turkey, UK, USA), however, we focus on Poland as a case study, since the country exhibits a clear bipartite structure dominated by two major parties. This allows us to make our argument in a basic case; further studies shall extend this work to other contexts. \autoref{tab:summary_table} summarize the main characteristics of our dataset.

\begin{table}[h]
    \centering
    \begin{tabular}{>{\raggedright}m{5cm}>{\centering\arraybackslash}m{3cm}}
        \hline
        \textbf{Category} & \textbf{Count} \\
        \hline
        Full User & 76507 \\
        Full Retweets & 232631 \\
        Full Influencer & 18483 \\
        Full Political Retweets & 7451 \\
        Full Politicians & 123 \\
        Full Political User & 3612 \\
        \hline
        Keyword Retweets & 64051 \\
        Keyword Influencer & 4290 \\
        Keyword User & 15584 \\
        Keyword Political Retweets & 3617 \\
        Keyword Politicians & 88 \\
        Keyword Political User & 2135 \\
    \end{tabular}
    \vspace{.2cm}
    \caption{Main characteristics of our dataset. We focus on retweets, influencer (retweeted accounts), politicians and users. We present these statistics for the full data (Full - \textit{Statistic}) and a sample obtained by filtering tweets with a set of keywords (Keywords - \textit{Statistic})}
    \label{tab:summary_table}
\end{table}

\subsection{Sampling techniques}
We consider the following sampling techniques.

\paragraph{Random sampling.}
We extract random subsets of the whole data, with varying sizes.

\paragraph{Keyword-based sampling.}
We create a list of keywords relating to general political terms in the language of the dataset, such as politics, election, vote, government, and prime minister. We include the names of the two main political parties in Poland at the time of collection (Law and Justice, far-right leaning, and Civic Platform, center-right leaning) as hashtags, to reduce the introduction of noise due to the widespread presence of such expressions in the language considered. We incorporate prominent figures from each political party and other well-known politicians during the considered period, trending political hashtags, and international political-related terms like `Trump', `Putin', `Biden', and `NATO'. We keep a tweet from the full dataset if it contains at least one of the selected keywords in its text. 

\paragraph{Seed-based sampling.} We choose a set of seed users, collect all of their tweets and the retweets of those. To this end, we consider the Twitter accounts of a set of politicians. We varied the size of this sample by selecting a fraction $\alpha$ of politicians with the most retweets in our subset.

\subsection{Latent Ideology Estimation}
To estimate the ideological positions of participants in the debate, we use a model originally proposed by~\cite{barbera2015birds, barbera2015tweeting}, adapting it in line with previous research such as ~\cite{flamino2023political, falkenberg2022growing,ramaciotti2022inferring}. 
This model leverages the assumption of homophily in social networks, which suggests that individuals tend to connect with others who are like-minded, particularly in political or ideological views. In this work, instead of focusing on follower/following dynamics, we utilize retweet interactions. A critical aspect of this approach is the selection of a subset of influencers, as their identification significantly impacts the results of the ideology estimation. We detail this selection process in the subsequent subsection. After identifying the influencer set, we apply the Correspondence Analysis technique \citep{greenacre2010correspondence}, comprising three main steps: (i) constructing the interaction matrix $A$, (ii) normalizing this matrix, and (iii) executing a singular value decomposition. Initially, we build matrix $A$, where $A_{ij}$ denotes the number of retweets from user $i$ to influencer $j$. Following this, we normalize $A$ by dividing each element by the total number of retweets:

\begin{equation}
P=\frac{A}{\sum_{ij} A_{ij}}.
\end{equation}

Next, we establish the following terms:

\begin{equation}
\begin{cases}
\textbf{r} = P \textbf{1}, \\
\textbf{c} = \textbf{1}^T P, \\
D_r = \text{diag}(\textbf{r}),\\
D_c = \text{diag}(\textbf{c}),
\end{cases}
\end{equation}

and proceed with the normalization operation:

\begin{equation}
S = D_r^{-1/2}(P- \textbf{r}\textbf{c}) D_c^{-1/2}.
\end{equation}

In the final step, we carry out a singular value decomposition $S= U \Sigma V^T$, where $U, V$ are orthogonal matrices and $\Sigma$ is a diagonal matrix of singular values of $S$. We determine the ideological orientation of users by examining the subspace associated with the primary singular value, assigning the ideological leaning of user $i$ based on the first column of $U$. Influencer ideology is then estimated by taking the median of their retweeters' ideology scores.

Note that the latent ideological space generated by this method cannot directly be interpreted as an opinion. However, the dimensions of this space present strong correlations with the left-right ideological axis in many European countries, as well as in the US, Canada, and Pakistan \citep{falkenberg2023affective,baqir2023social}. Furthermore, these latent spaces can be mapped to real-life ideological spaces on several topics, such as immigration and climate change, among others \citep{ramaciotti2022inferring,peralta2024multidimensional}.

\subsection{Comparison metrics}
We evaluate the level of polarization on the distribution of latent ideologies and compare different samples. We first normalize the ideologies so they have mean 0 and standard deviation 1. We measure the level of polarization in the users' ideology distribution in terms of multimodality of the distribution through the Dip Test and Statistic.
Moreover, we compare different ideology distributions through the Wasserstein Distance. 
Finally, we evaluate some graph measures on the retweet network.

\paragraph{Dip test and statistic.} We quantify polarization in terms of multimodality using Hartigan’s diptest. 
Hartigan’s diptest is a nonparametric test to measure the multimodality of a distribution from a sample \citep{hartigan1985dip}. It calculates the maximum difference over all sample points between the empirical distribution function and the unimodal distribution function that minimizes such maximum difference. The test produces a statistic $D$, which quantifies the magnitude of multimodality, and a statistical significance $p$. If $p<0.01$, we say that the ideology distribution shows statistically significant multimodality. Conversely, if $p \geq 0.01$, we cannot reject the unimodality of the distribution.

\paragraph{Wasserstein distance.} The Wasserstein distance, also known as the Earth Mover's distance, quantifies the minimal cost required to transform one distribution into another, considering the geometry of the underlying space. It is defined as 
\begin{equation}
    W(P, Q) = \inf_{\gamma \in \Gamma(P, Q)} \mathbb{E}_{(x,y) \sim \gamma}[d(x, y)],
\end{equation}
where \(\Gamma(P, Q)\) represents the set of all joint distributions with marginals \(P\) and \(Q\). The Wasserstein Distance provides a meaningful measure even when the distributions do not overlap, and the closer the distance to 0, the more similar the two distributions are. 


\paragraph{Graph Measures}
The measure we adopted to detect the network structure is the relative size of the largest weakly connected component (LWCC). 
The LWCC refers to the largest subset of nodes where every node is reachable from any other node. It is possible considering the network's directed edges without considering edge directions. LWCC is used to understand overall network connectivity and structural resilience. Mathematically, the size of the largest weakly connected component is defined as: 
    
\begin{equation}
   LWCC = \max_{S \in \text{weakly connected components}} |S|
\end{equation}

We use the NetworkX package implemented in Python to calculate network metrics component \citep{hagberg2008exploring}.

\section{Results}

\subsection{The role of the political debate in Twitter discussion}

Twitter has been often used as a source to study political debates \citep{falkenberg2022growing,falkenberg2023affective,peralta2024multidimensional,cinelli2021echo,garimella2017longterm,garimella2018political,williams2015network}. Yet, it is not clear how central the political discussions are to the wider Twitter discussion. In the United States, \citet{wojcieszak2022most} show that most Twitter users are not politically engaged, but those that are show strong patterns of political homophily. However, whether these observations extend to other countries is not clear. Hence, we start our analysis by studying the polarization of the Polish political debate and how they relate to non political interactions.

\begin{figure}[ht!]
     \centering
     \begin{subfigure}[t]{0.45\textwidth}
         \centering
         \includegraphics[width=\textwidth]{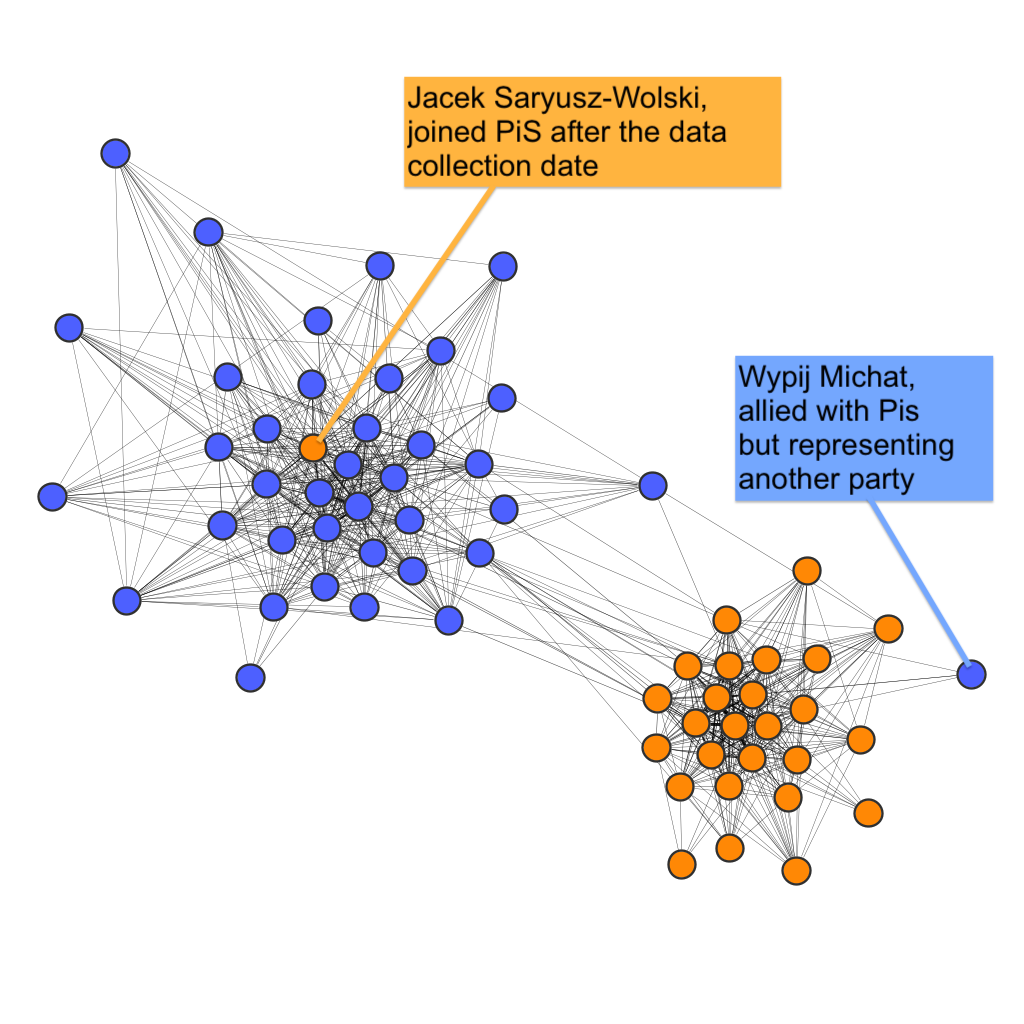}
     \end{subfigure}\hspace{1cm}
    \begin{subfigure}[t]{0.45\textwidth}
         \centering         \includegraphics[width=\textwidth]{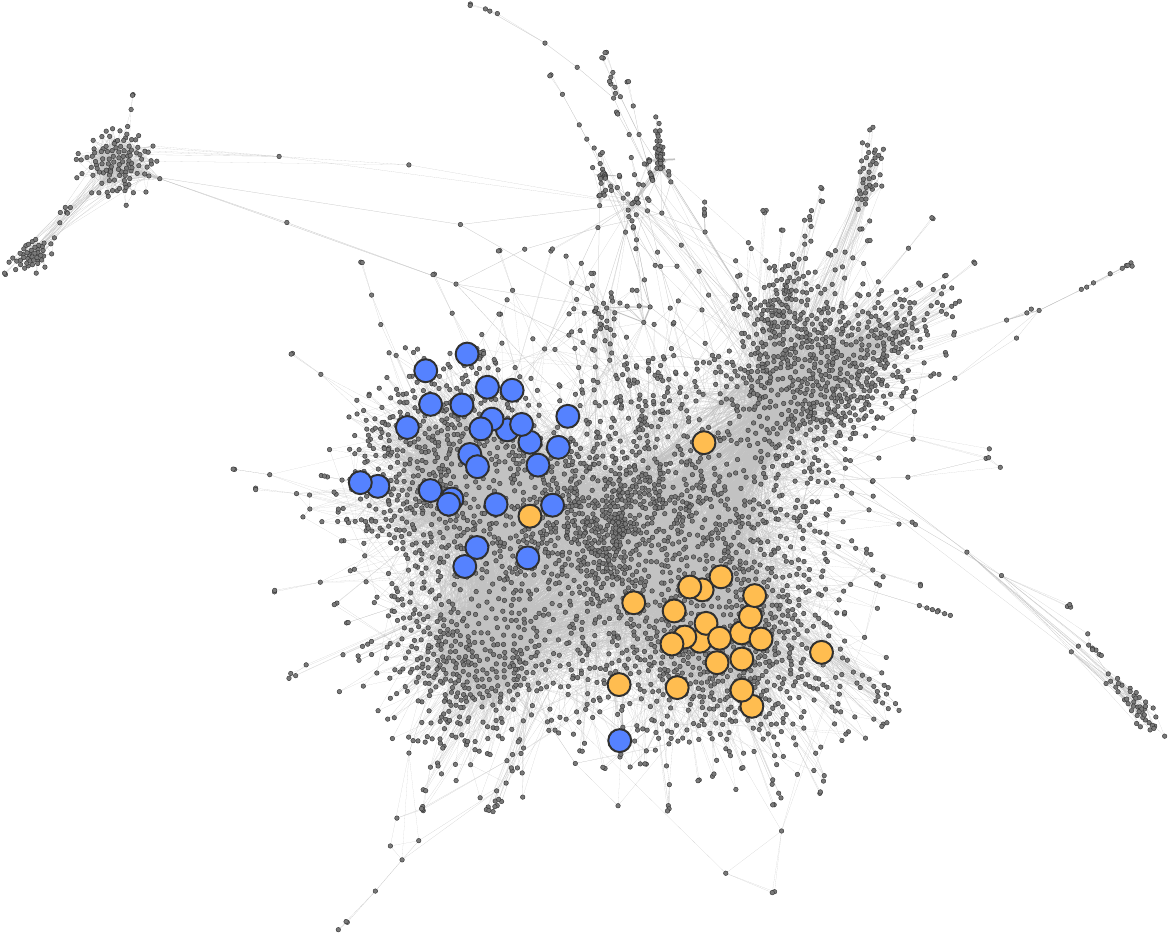}
     \end{subfigure}
    \caption{\textbf{Visualization of the Twitter co-occurrence network in Poland.} Left: politicians only. Right: politicians and non-politicians. Blue circle are the PiS politicians, while orange circles correspond to PO politicians. The left section displays the co-occurrence network of the political debate in Polish tweets. This plot shows a clear polarization between the network nodes, with distinct political polarization. On the right side, the co-occurrence network of the entire debate in Polish tweets is shown. In this case, the polarization observed in the left section is not as evident, as it is merged within the broader context of the overall debate in Poland.}
    \label{fig:co_occurancy}
\end{figure}

To understand the role of the political debate in the wider Twitter discussion, two retweet co-occurrence networks are presented in \autoref{fig:co_occurancy}. The left section displays the co-occurrence network of Polish politicians. This plot shows a clear polarization between the network nodes, with distinct political orientations. On the right side, the co-occurrence network of the entire debate in Polish tweets is shown. In this case, the polarization observed in the left section is not as evident, as it is merged within the broader context of the overall debate in Poland. In both cases, politicians are sorted in a clear community structure, that corresponds with party affiliations almost perfectly. Outliers include politicians that defected and changed parties after the affiliation data was collected. However, if political content is not carefully separated from apolitical content, the identification of polarized social network structures may be obscured by other content.

\subsection{Reconstructing polarization with partial information}

\begin{figure}[!t]
     \centering
     \begin{subfigure}[b]{0.45\linewidth}
        \centering
        \includegraphics[width=\textwidth]{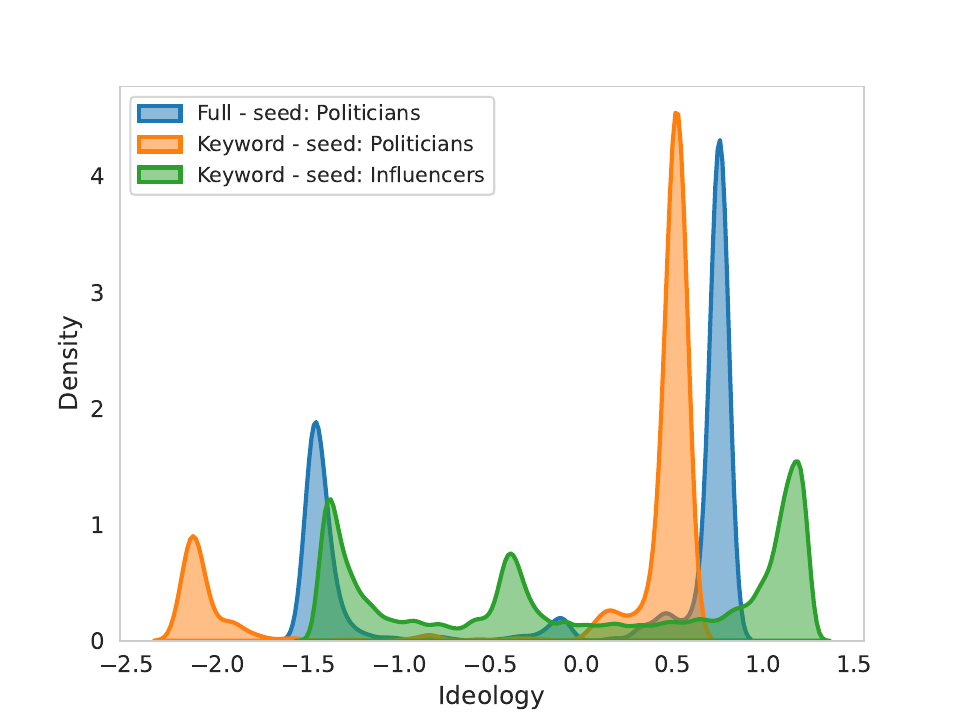}
        \caption{}
        \label{fig:three_methods}
     \end{subfigure}
   \quad
 \begin{subfigure}[b]{0.45\linewidth}
        \centering
        \begin{tabular}{>{\raggedright}m{2cm}>{\raggedright}m{2cm}>{\centering\arraybackslash}m{1cm}>{\centering\arraybackslash}m{1cm}>{\centering\arraybackslash}m{1cm}}
            \hline
            \textbf{Data} & \textbf{Retweets} & \textbf{Users} & \textbf{Influencers} \\
            \hline
            Full-Politicians &   7,451 & 3,612 & 123 \\
            Keyword-Politicians &   3,617 & 2,135 & 88 \\
            Keyword-Influencers &  64,051 & 15,584 & 4,290 \\
            \hline
        \end{tabular}
        \vphantom{\includegraphics[height=1.5em]{figures/three_methods.pdf}}
        \caption{Summary of Retweets, Users, and Influencers}
        \label{tab:fig_three-methods}
    \end{subfigure}
  \caption{\textbf{Comparison of ideology distributions obtained across different sampling strategies using the full 24-hour Poland dataset.} Left: The figure shows distributions for: a) the full dataset with politicians as seeds for latent ideology estimation (blue), b) the dataset obtained through keyword- and seed-based sampling with politicians as seeds (orange) and c) uses the same strategy as b) but using all users as influencers in the latent ideology estimation (green). The right panel presents basic statistics for each distribution. The plot indicates that a) and b) result in bimodal distributions. Looking at the dimensions and considering the used strategies, b) is a partial sample of a), and thus keyword-based sampling is underrepresenting one of the peaks. In contrast, c) shows that when not limited to politicians as seeds, a third peak of moderate users is captured, resulting in a less polarized distribution. }
  \label{fig:three_methods_panel}
\end{figure}

\begin{figure}[!t]
     \centering
    \includegraphics[width=\textwidth]{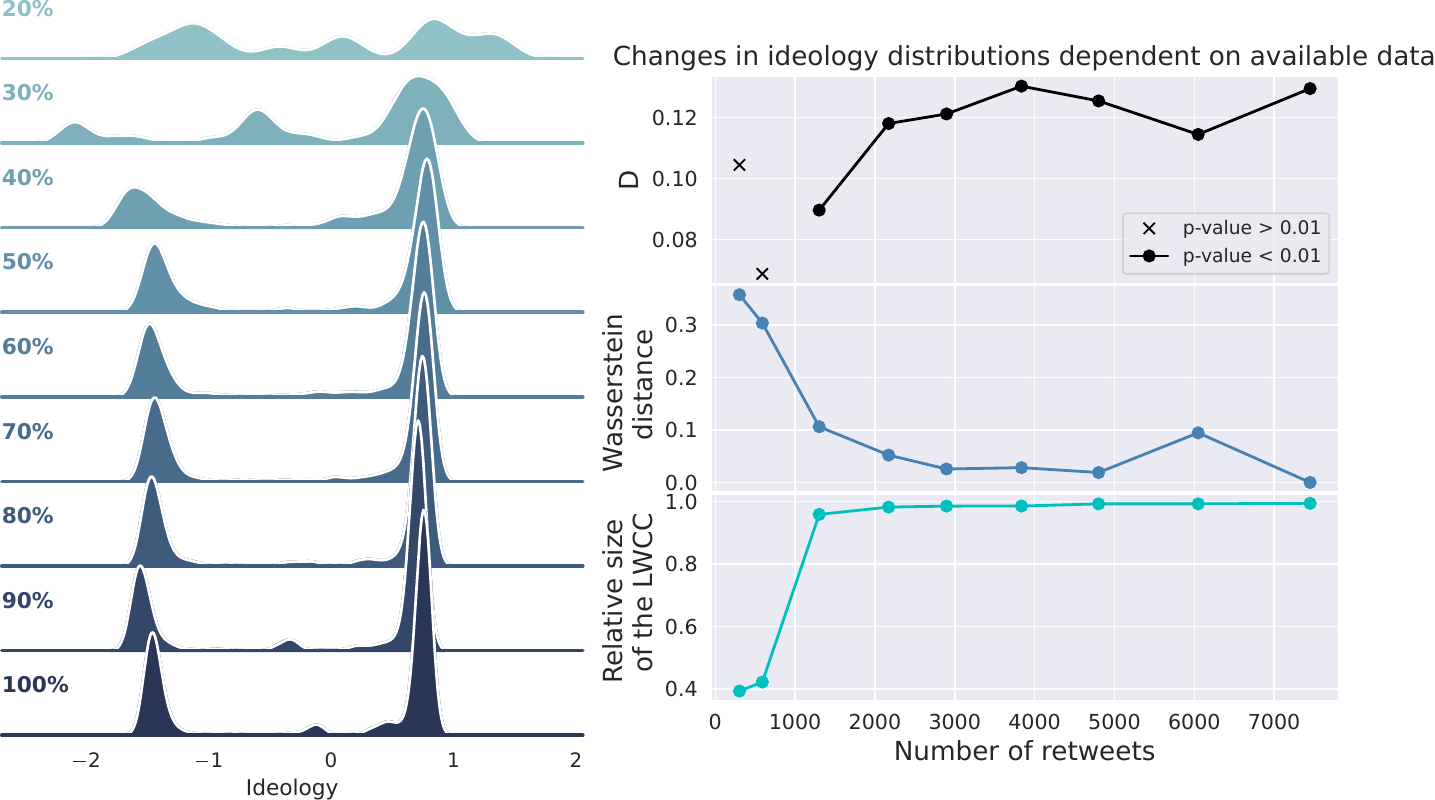}
    \caption{\textbf{Study of varying the percentage of accessed data using the Polish 24h dataset with politicians as seeds for latent ideology retrieval.} Left: Ideology distributions for various percentages of partial data. The bimodal distribution collapses for percentages below 30\%. Right: Corresponding metrics for each percentage (x-axis represents the number of retweets obtained). In order: Hartigan’s diptest with significance levels, Wasserstein distance to the 100\% distribution, and relative size of the Largest Weakly Connected Component (LWCC). For 20\% and 30\% samples, multi-modality is not significant, with the bimodality statistic (D) increasing as the data percentage rises. The Wasserstein distance drops sharply for initial samples and stabilizes above 50\% (around 2,000 retweets). The relative LWCC size indicates a dismantelling of the retweet network for the first two sub-samples.}
    \label{fig:panel1_partial}
\end{figure}

\begin{figure}[!t]
     \centering
    \includegraphics[width=\textwidth]{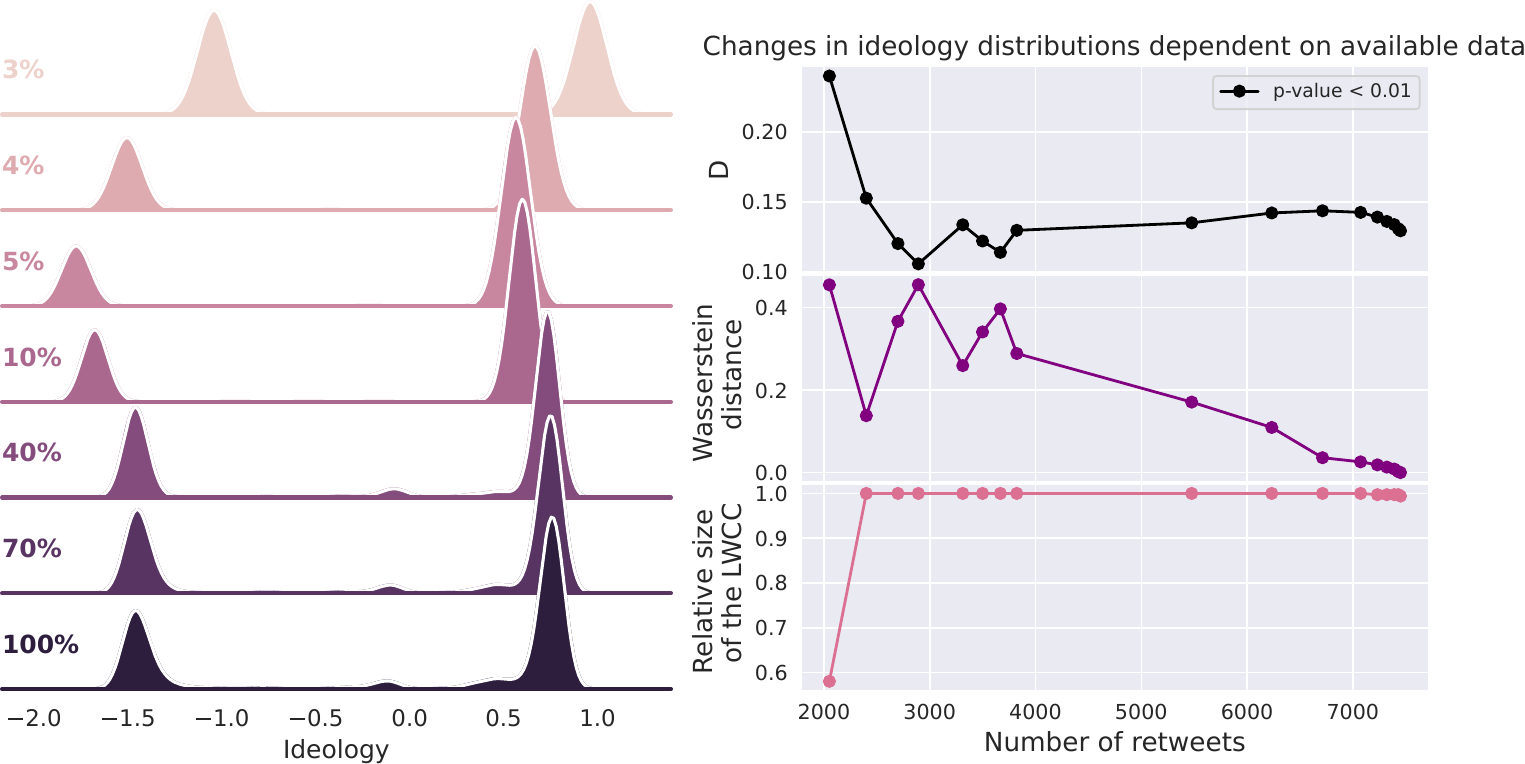}
    \caption{\textbf{Study of varying the percentage of top retweeted politicians as seeds using the Polish 24h dataset for latent ideology retrieval}. Left: This figure shows ideology distributions for different percentages of seed politicians, revealing bimodality across all percentages but with variations below 40\% of top politicians. Lower percentages than 3\% do not produce viable results. Right: corresponding metrics for each percentage, with the x-axis representing the number of retweets obtained at different percentage levels. The metrics include Hartigan’s diptest with significance levels, Wasserstein distance to the 100\% distribution, and the relative size of the Largest Weakly Connected Component (LWCC). For samples with fewer than 4,000 retweets, both the bimodality statistic (D) and the Wasserstein distance fluctuate. As the number of retweets increases to 4000 and higher, the bimodality statistic stabilizes around 0.15, and the Wasserstein distance decreases to 0. The relative size of the LWCC indicates that the network is only disconnected for the top 3\% of influencers and less.}
    \label{fig:panel2_partial}
\end{figure}

We continue our analysis by studying how the amount of data changes the measured level of polarization extracted from structurally polarized retweet networks. 
In particular, we consider the following samples from the complete dataset:
\begin{enumerate}
    \item Full data with all politicians as seeds (\autoref{fig:three_methods_panel});
    \item Keyword-based sampled data with all politicians as seeds (\autoref{fig:three_methods_panel});
    \item Random sample of full data with varying sizes, using all politicians as seeds (\autoref{fig:panel1_partial});
    \item Keyword-based sampled data with all users as seed (\autoref{fig:three_methods_panel});
    \item Full data with varying percentages of politicians as seeds (\autoref{fig:panel2_partial}).
\end{enumerate}

From the sampled data, we dropped all users that only retweeted a single user, since we cannot infer the relative positions of the users if they only retweeted one other user (see Methods).

Below, we outline the three main results of this section. We start by first comparing three different strategies to sample data and construct the retweet network used for the latent ideology estimation assuming full access to the full 24-hour Poland dataset (\autoref{fig:three_methods_panel}). Then, we look into more detail the obtained distributions when we have partial access to the dataset for one of these cases (\autoref{fig:panel1_partial}), in which we use the full list of politicians as seeds. Finally, we assess the effects of varying the number of politicians as seeds (\autoref{fig:panel2_partial}).

The overall ideology distribution of different sampling methods in \autoref{fig:three_methods_panel} shows very similar distributions for using the full data with politicians as seeds and the keyword sampled data with politicians as seeds. However, the keyword sampling does not sub-sample the conversation uniformly and seems to introduce a bias, in which one of the sides is less prominent. 
Moreover, when we consider all users as influencers, instead of only politicians, in the keyword-based sample, a new moderate peak shows up and we find a less polarized distribution. 
As we have seen in \autoref{fig:co_occurancy} when we extend the analysis to a broader context, the political polarization is less evident.
Interestingly, even though we use political keywords in the keyword-based sampling, the level of polarization differs if we consider only politicians or all users as influencers.

When using random sub-samples of the full data and then using all political accounts as seeds (\autoref{fig:panel1_partial}) we find that up until 40\% there is not enough data to provide accurate estimates of the polarized structure of the conversation network. For these values, we cannot reject the unimodality of the distribution. From that point onward, the obtained distributions are bimodal and stabilize to larger values of the bimodality statistic (D) with increasing amounts of data. Similarly, the distance between each obtained distribution and the distribution obtained from all data sampled from political seeds (100\%), decreases with the amount of data used. 
We find that, for the percentages lower than roughly 40\%, the non-significance in the D Statistic and the very large values of distributional distance strongly correlate with a breakage of the network into smaller largest weakly connected components of the underlying retweet network. 

Conversely, when maintaining the full dataset but varying the number of politicians used as seeds (\autoref{fig:panel2_partial}), we find that, even though bimodality is maintained, results are unstable for lower percentages. Results show that the latent ideology produces consistent polarisation measures if the number of politicians used as influencers is larger than \~20\% of the top retweeted politicians (in this case, corresponding to 25 politicians). From this percentage onward, we find consistent values of the D statistic and a slightly decreasing distance compared to the case of using 100\% of the politicians. We also find that for the 3\% case, the network is dismantled and thus the distribution is not representative even if yielding significant results for the D statistic. 

We find that we can approximate the level of polarization present in our data with limited sample sizes. However, smaller sizes might become problematic, especially for cases in which the retweet network is not complete anymore and breaks into smaller components. Additionally, different methods reflect the overall level of polarization to different extents. We find that when using random samples with all politicians as seeds, we can approximate the level of polarization with less data than when using the full data, but with varying sizes of the political seeds. Therefore, it is crucial to use a large enough sample of relevant influencers. Furthermore, we find that keyword-based filtering might introduce some bias and over/underrepresent one side of the conversation. While the ideology distribution still reflects similar positions of polarization, we find peaks of different sizes, which indicates that using keywords might favor sampling the activity from one of the sides. Therefore, the choice of keywords is not trivial and researchers should take into account alternative methods to validate their choices.

\subsection{Keywords representativeness}
Finally, we investigate whether different choices on the set of keywords used to download the data can impact the results of the analysis. We compare the ideology distribution of the full-seed Politicians dataset (considered as baseline) with the distribution obtained by selecting tweets containing keywords related to the Polish political debate. The author ideology we used in both cases refers to the one of the baseline dataset.

\begin{figure}[!t]
    \centering
    \begin{subfigure}[b]{0.48\textwidth}
        \centering
        \includegraphics[width=\textwidth]{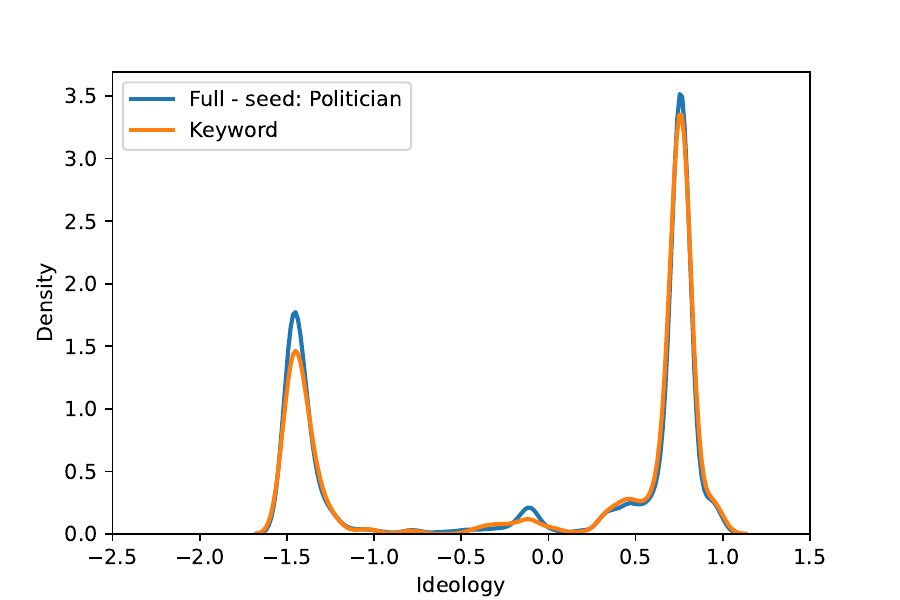}
        \caption{}
        \label{fig:full_keyw}
    \end{subfigure}~
    \begin{subfigure}[b]{0.48\textwidth}
        \centering
        \includegraphics[width=\textwidth]{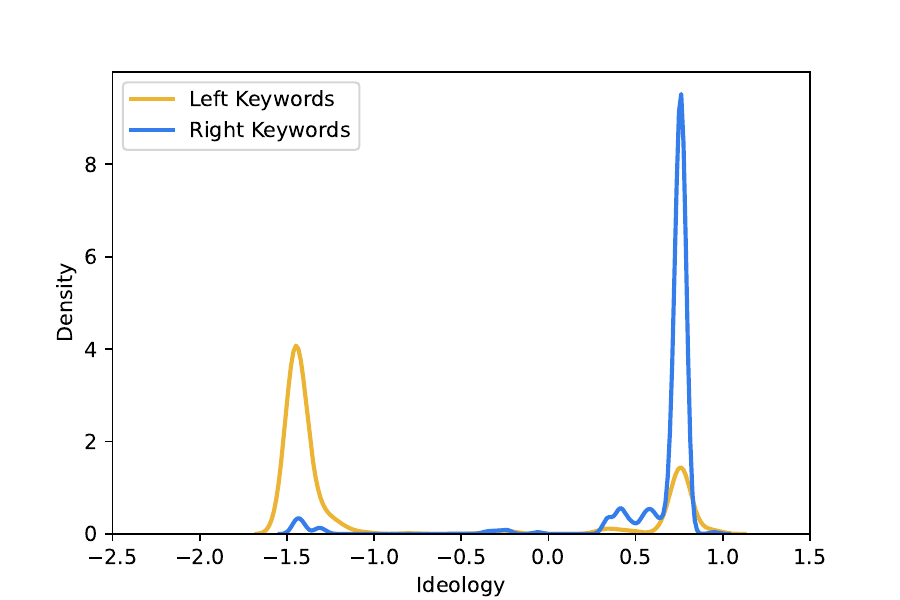}
        \caption{}
        \label{fig:left_right}
    \end{subfigure}
    \caption{\textbf{Keyword based samples can substantially bias the inferred distribution of ideological opinions on Twitter.} Left: Using a broad range of political terms can closely approximate the ideology distribution computed using the full dataset and politicians as seed nodes. Right: Poorly selected keyword samples can result in significant bias in the identified ideology distribution, resulting in the overrepresentation of either the political left (yellow) or the political right (blue).
    }
    \label{fig:network_comparison}
\end{figure}

As illustrated in \autoref{fig:full_keyw}, the two distributions are quite similar, indicating that the keyword-based selection approximates the baseline distribution well. This demonstrates that a set of carefully selected keywords can effectively capture the ideological distribution of the full dataset.

\begin{figure}[!t]
    \centering
        \includegraphics[width=0.5\textwidth]{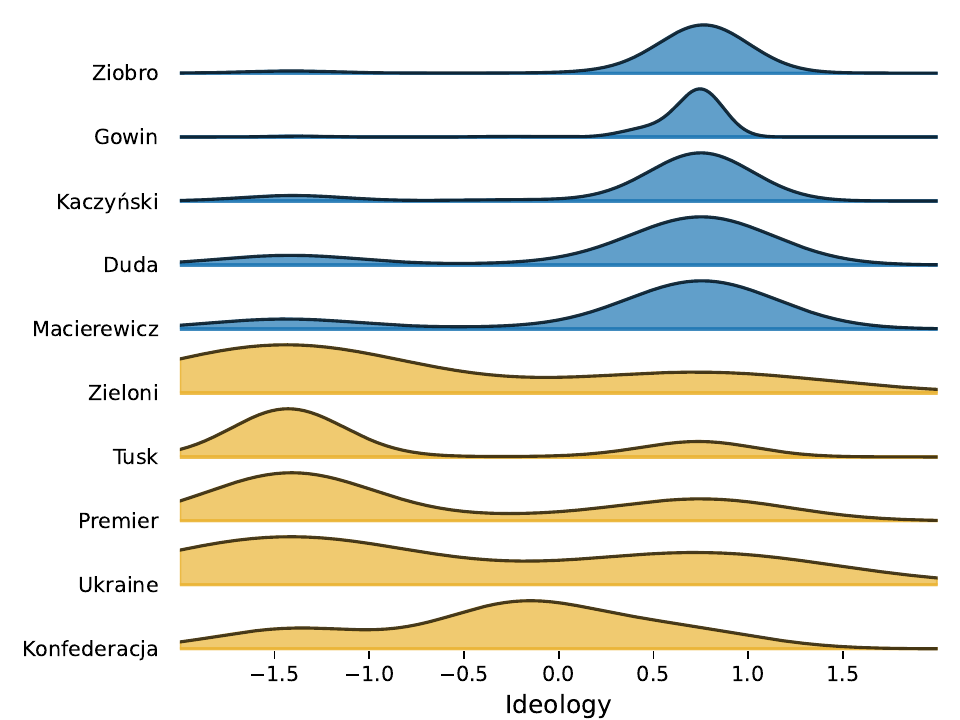}
        \label{fig:subfig1}
    \caption{\textbf{A selection of keywords which exhibit strong ideological bias towards either the political left or the political right.} Each distribution shown corresponds to the distribution of Twitter users' ideology scores who used the keyword in question. The distributions in blue correspond primarily to terms used by the political right. The distributions in yellow correspond to words used primarily by the political left.}
    \label{fig:words_lf_rt}
\end{figure}

However, it is crucial to consider that if the keywords do not encompass the full spectrum of ideologies relevant to the topic, the resulting data may be unbalanced and the result biased. To explore this, we divided our keyword dataset into two subsets based on the mean ideology of the associated tweets, representing the two ends of the political spectrum.
\autoref{fig:left_right} and \autoref{fig:words_lf_rt} illustrate how keywords are not used equally by the two factions, but opposite ideologies tend to use disjoint sets of keywords. Specifically, \autoref{fig:left_right} shows a significant shift in ideology distribution if only left- or right-leaning keywords are used, highlighting the possible introduction of biases in case of inaccurate keywords selection. Furthermore, \autoref{fig:words_lf_rt} displays the ideological distribution linked to specific keywords, emphasizing how certain keywords correlate with distinct ideological perspectives.

Hence, while a keyword-based selection can approximate the overall ideological distribution, it is essential to carefully select the keywords used for collection to avoid the introduction of biases conflating the results.

\section{Discussion}
In this paper, we analyzed the impact of data sampling on the identification of polarized social network structures on Twitter. Using a dataset covering 24h of discussion in Poland, we first analyzed the presence of political polarization and discussed its relationship with the non-political content of the debate. We then turned our attention to the impact of data sampling in the assessment of polarization, and highlighted the possible presence of biases in case of data retrieval based on a set of keywords. Our results suggest numerous ways in which the study of polarized social network structures can be biased when using a sampled dataset as opposed to a complete dataset. 

First, we showed that the political discussion on Twitter is only a subset of the wider Twitter discussion, such that the identification of political polarization requires a careful approach to disentangling political from apolitical content online. This aligns with previous research in the United States which found that most Twitter users are not politically aligned \citep{wojcieszak2022most}. Second, we use the latent ideology method to extract a synthetic ideology distribution from the identified political interaction network on Twitter, but where the political interaction network is extracted using different methods. From this synthetic ideology distribution, polarization can be quantified through a variety of methods. We show that extracting the latent ideology by focusing on interactions with politicians, and by using keyword based sampling using a broad range of political keywords, both result in the identification of a multimodal ideology distribution. However, we show that the modality of ideology distribution can change depending on whether the latent ideology is computed focusing on politicians, or focusing on popular accounts in general, even if tweets are sampled to only include political keywords.

Third, we consider the ideology distributions computed using politicians as the seed for the latent ideology and then down-sample our full political communication network to see how the modality of the ideology distribution changes as a function of the sampling rate. We show that for sampling sizes above $40\%$, the ideology distribution is reasonably stable, as quantified using Hartigan's diptest for multimodality and the Wasserstein distance to the full ideology distribution. However, below $40\%$ sampling rates, we see that the identification of a multimodal ideology distribution is no longer statistically significant. However, if the full dataset is available, we show that the ideology distribution computed is relatively stable, and the computed multimodality is statistically significant, even if only a very small number of politicians are selected to compute the latent ideology. These results emphasise that it is unlikely that polarization can be fairly measured on social media using small sample datasets, and that researchers must have access to comprehensive datasets in order to ensure robust results. 

Finally, we looked in greater detail at keyword-based sampling. This is particularly important since, historically, the majority of studies on social media have acquired data using a small number of seed keywords to identify relevant content. Our results show that if the set of keywords is chosen carefully, the identified ideology distribution closely approximates the ideology distribution computed using the full dataset with politicians as seed nodes. However, we show that several keywords are ideologically biased and are used primarily by the political left or by the political right. If these keywords are used to sample the full dataset, the resulting ideology distribution can exhibit significant political bias. This emphasises that great care must be taken when inferring political ideologies online based on keyword datasets.

Our work has important limitations which should be considered in future work extending this analysis. Our analysis focuses on polarization in the Polish Twittersphere, and could be expanded to other social media platforms and countries. There is also an open question on how polarization can be measured on social media. Our approach is based on the existing literature, specifically the works of \cite{barbera2015birds} and \cite{falkenberg2022growing}, which effectively measure polarization using latent ideology. However, there may be other methods to measure polarization.


Our results highlight the importance of quality data to conduct research with social media. Social media platforms have a profound impact on our societies, from politics \citep{vishnuprasad2024tracking,papasavva2024comprehensive,bouchaud2024meta,kaiser2019targeted,lelkes2017hostile,theocharis2016does}, to news consumption \citep{shearer2024how,allcott2020welfare}, health \citep{cinelli2020covid,monsted2022characterizing,micallef2020role} and climate \citep{falkenberg2022growing,williams2015network,torricelli2023does}, among others. As these platforms have played a key role in the interference of democratic process \citep{kaiser2019targeted,bouchaud2024meta}, it is especially concerning that they are not easily accessible to academics for research purposes. 

New opportunities such as the European Union’s Digital Services Act (DSA) are a first step that might ensure data access for academic and other public research institutions. Initial findings indicate that compliance with the DSA by technology companies has been sub-optimal \citep{jaursch2024enabling,bouchaud2024meta}, highlighting the necessity for more robust research on the implications of the DSA and the ramifications of its practical implementation. Nevertheless, without a precise understanding of which data are needed to reliably study different aspect of online social media, such efforts may be vain, and platforms can be compliant with legislation by providing access to few data that are of little utility in practice. In this regard, our work can contribute to the design of technical guidelines for legislators and ensure that quality data are provided by the platforms to researchers. Finally, our work contributes to shed light on the reliability of social media studies which may be biased if their datasets have been acquired through ill thought out sampling strategies.

\section{Acknowledgements}
This work is the output of the Complexity72h workshop, held at the Universidad Carlos III de Madrid in Leganés, Spain, 24-28 June 2024. \url{https://www.complexity72h.com}.

\bibliographystyle{abbrvnat}
\bibliography{references} 

\end{document}